\documentclass[12pt, dvips]{article}
\usepackage{epsfig}

\begin{document}

\newcommand{\p}{\partial}
\newcommand{\om}{\omega}
\newcommand{\e}{\epsilon}
\renewcommand{\a}{\alpha}
\renewcommand{\b}{\beta}

\title{Escape Probability and Mean Residence Time 
       in Random Flows with Unsteady Drift
\footnote{This work was begun during a
visit to the Oberwolfach  Mathematical  Research
Institute, Germany.  }   }
	
\author{James R. Brannan, Jinqiao Duan and  Vincent J. Ervin \\
 \\
 Clemson University \\
Department of Mathematical Sciences \\
Clemson, South Carolina 29634, USA. \\
E-mail: {\em duan@math.clemson.edu}   }

\date{September 1, 1999 }

\maketitle

\begin{abstract}

We investigate fluid transport  in random velocity fields
with unsteady drift. First, we propose to quantify
fluid transport between flow
regimes of different characteristic motion, by escape probability and mean
residence time.
We then develop numerical  algorithms   to solve for
escape probability and mean residence time,
 which are described by 
backward  Fokker-Planck type partial differential equations.
A few computational issues are also discussed.  Finally,  
we apply these
ideas and numerical algorithms      to 
a tidal flow model.
 

\bigskip
{\bf Short Running Title:}   Transport in Random Flows

\end{abstract}

\newpage
\section{Introduction}

The Lagrangian view of fluid motion is particularly important in geophysical
flows since only Lagrangian data can be obtained in many situations. 
It is essential to understand fluid particle trajectories
in many fluid problems.
  
Stochastic dynamical systems arise as models for
fluid particle motion  in geophysical flows with
random velocity field 
\begin{eqnarray}
 \dot{x} & = & f(x,y, t)   +  a(x,y)  \dot{w}_1, \label{eqn1} \\ 
 \dot{y} & = & g(x,y, t)   +  b(x,y)  \dot{w}_2, \label{eqn2}  
\end{eqnarray}
where  $w_1(t), w_2(t)$  
are two real independent Brownian motion processes,
$f, g$ are deterministic drift part,  
and $a, b$ are the intensity coefficients of   diffusive noise part,
of the velocity field. Note that the  generalized derivative of
a Brownian motion process is a mathematical model for
``white noise". For general background in stochastic dynamical systems,
see \cite{Arnold2, Freidlin, Oksendal}.

Deterministic quantities, such as escape probability (from a fluid domain)
and mean residence time (in a fluid domain),
that characterize stochastic dynamics can be computed
by solving the Fokker-Planck type backward partial differential equations.

In a previous paper \cite{Brannan-exit}, when the drift is steady, i.e, 
$f, g$ do not depend on time,
we have  quantified
fluid transport between flow
regimes of different characteristic motion by
escape probability and mean
residence time;
developed methods for computing 
escape probability and mean exit time; and applied these methods
in the investigation of geophysical fluid dynamics.   
In this paper, we further consider the case of unsteady
or nonautonomous drift $f(x,y,t), g(x,y,t)$, develop
a numerical algorithm  for computing escape probability and mean exit time,
and demonstrate the application of this approach to a tidal flow model.

\section{Stochastic Dynamics \\of Fluid Particle Motion }

For a planar bounded domain $D$, we can consider the exit problem
of random solution trajectories of (\ref{eqn1})-(\ref{eqn2}) 
from $D$. To this end, let $\p D$ denote the boundary of $D$.
The residence time of a particle initially at $(x,y)$ at time $t$
inside $D$
is the time until the particle first hits $\p D$ (or escapes from $D$).
The mean residence time $\tau(x,y,t)  = t + u(x,y,t)$,
where $u(x,y,t)$  satisfies \cite{Hasminskii}
\begin{eqnarray} 
 u_t +   \frac12 a^2 u_{xx} + \frac12 b^2 u_{yy} & + & f(x,y,t) u_x
               + g(x,y,t) u_y    =    -1,   \label{eqn3}\\
  u(x,y,t) & = & 0, \; (x,y,t) \in \p D \times (0, T),  \label{eqn7} \\
  u(x,y, T) & = & 0, \; (x,y) \in  D,  
\end{eqnarray}  
where $T$ is big enough, i.e., $T > \sup \tau(x,y,t)$ and here the
$\sup$ is taken over all $(x, y)$ in a compact set,
i.e., the closure of the domain $D$.

Note that $u(x,y,t) = \tau(x,y,t) - t $ is the mean residence time
after time instant $t$, or it quantifies how much longer
a particle will stay inside $D$ after we observe it at position
$(x,y)$ at the time instant $t$.

Let $\Gamma$ be a part of the boundary $\p D$.
The escape probability $p(x,y,t)$ is the probability   
that the trajectory of a particle starting at position $(x,y)$ and
at instant $t$ in $D$
first hits $\p D$ (or escapes from $D$) at some point in $\Gamma$
(prior to escape through $\p D - \Gamma$), and 
$p(x,y, t)$ satisfies \cite{Lin, Schuss} 
\begin{eqnarray} 
 p_t +  \frac12 a^2 p_{xx} + \frac12 b^2 p_{yy}  &  +  &f(x,y,t) p_x
               +g(x,y,t) p_y   =   0,  \label{eqn4} \\  
 p(x,y,t) & = & 0, \; (x,y,t) \in (\p D-\Gamma ) \times (0, T), \\
 p(x,y,t) & = & 1, \; (x,y,t) \in  \Gamma \times (0, T), \\
 p(x,y, T) & = & 1, \; (x,y) \in  D. 
\end{eqnarray} 
Note that $p(x,y,t)$  depends on  $\Gamma$, and it may be better denoted as
$p_{\Gamma}(x,y,t)$.

Suppose that initial conditions (or initial particles) are 
uniformly distributed over $D$. 
The average escape probability $ P(t) $ that a trajectory
will leave $D$ along the subboundary $\Gamma$ at time $t$, before leaving the
rest of the boundary, is given by  \cite{Lin} 
\begin{equation}
	  P(t) =  \frac{1}{|D|} \int\int_D p(x,y,t) dxdy,
	  \label{average}
\end{equation} 
where $|D|$ is the area of domain $D$.

\section{Numerical Approaches}

For the  backward type of partial differential equation (\ref{eqn3}),
we reverse the time
$$
	s= T-t
$$
Then the mean residence time $u(x,y,s)$ (we still use the same notation)
 satisfies
\begin{eqnarray} 
 u_s & = &  \frac12 a^2  u_{xx} + \frac12 b^2 u_{yy}  \nonumber \\
     & + &  f(x,y,T-s) u_x
               + g(x,y,T-s) u_y  +  1,   \label{new1}\\
  u(x,y,s)  & = & 0, \; (x,y,s) \in \p D  \times (0, T),  \label{new2} \\
  u(x,y,0)  & = & 0, \; (x,y) \in  D.    \label{new3}
\end{eqnarray}

Similarly for the escape probability $p(x,y,s)$ 
(we still use the same notation),  we have     
\begin{eqnarray} 
 p_s &=&  \frac12 a^2 p_{xx} +\frac12 b^2 p_{yy}  + f(x,y,T-s) p_x
               +g(x,y,T-s) p_y ,  \label{new4} \\  
 p(x,y,s) & = & 0, \; (x,y,s) \in (\p D-\Gamma ) \times (0, T), \label{new5}\\
 p(x,y,s) & = & 1, \; (x,y,s) \in  \Gamma \times (0,T),    \label{new6}\\
 p(x,y, 0) & = & 1, \; (x,y) \in  D. 	\label{new7}
\end{eqnarray}

A piecewise linear, finite element approximation 
scheme \cite{Brenner} was used
for the numerical solutions of the escape probability $p(x,y, s)$,
and the mean residence time $u(x,y, s)$, described by the parabolic
equations (\ref{new1}), and (\ref{new4}), respectively.  
By transforming back to original time
$t=T-s$, we get $p(x,y,t)$,
and   $u(x,y, t)$. We have used a few different
time-discretization schemes, including  the implicit backward
in time, and Crank-Nicholson scheme \cite{Eriksson}.
The code works also for boundary defined by  a 
collection of points lying on the boundary. A piecewise cubic
splines were constructed to define such boundaries.

\section{Application to a Tidal Flow}
 
To demonstrate the above ideas and numerical algorithms,
we consider a tidal flow model. This flow model is very idealistic
and here we just use it as an illuminating example.
 
Oscillatory tidal water motions dominate a large part of the
coastal regions.  Beerens and Zimmerman \cite{Beerens_Zimmerman}
considered      a   tidal flow model with velocity field   
 
\begin{eqnarray}
u_o &=& \pi  \sin(\pi x)\cos(\pi y)  + \pi  \lambda \cos(2\pi t), 
	\label{tidal1}  \\
v_o &=& -\pi \cos(\pi x)\sin(\pi y),  \label{tidal2}
\end{eqnarray}
where  $\lambda$ is a parameter measuring the intensity of the tidal
wave $\pi \cos(2\pi t)$. We take $0< \lambda < 3$ as used
by Beerens and Zimmerman \cite{Beerens_Zimmerman}.

As pointed out by Beerens and Zimmerman \cite{Beerens_Zimmerman},
it is essential to include more complicated temporal modes in this model
in order to describe more realistic tidal flows. We include random 
temporal modes or white noise in this model, i.e., we consider a tidal 
flow model with unsteady drift part and random diffusive part:

\begin{eqnarray}
\hat{u}_o &=& \pi  \sin(\pi x)\cos(\pi y)  + \pi  \lambda \cos(2\pi t)
		+ \sqrt{2\e} \dot{w}_1, \label{tidal3}\\
\hat{v}_o &=& -\pi \cos(\pi x)\sin(\pi y) + \sqrt{2\e} \dot{w}_2 ,\label{tidal4}
\end{eqnarray}
where $\e >0$ is the constant intensity of the white noise. We assume that
the random temporal modes are weaker than the time-periodic mode
and so we take $0<\e<0.1$ in the following simulations.

\begin{figure}[tbp] 
\hbox to \hsize{\hfil \psfig{figure=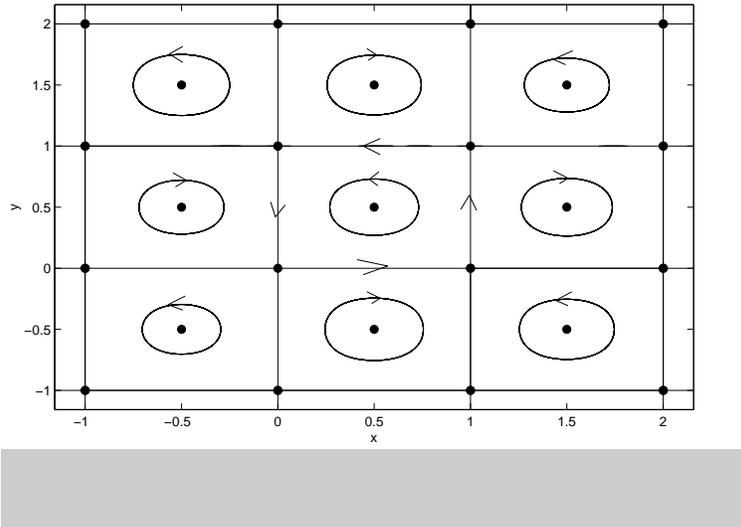,height=3in}\hfil}
\caption{Unperturbed flow}
\label{cellular}
\end{figure}

We study the transport of fluid particles 
in this tidal flow model. The equations of motion of fluid particles
in this tidal flow are
\begin{eqnarray}
\dot{x} & = &  \pi  \sin(\pi x)\cos(\pi y)
	 +\pi \lambda \cos(2\pi t)           + \sqrt{2\e} \dot{w}_1, \\ 
 \dot{y} & = & -\pi \cos(\pi x)\sin(\pi y)   + \sqrt{2\e} \dot{w}_2,  
\end{eqnarray}
where $w_1(t), w_2(t)$  are two   independent Brownian motion processes.
The unperturbed flow, with no temporal periodic ``tidal"     mode 
$ \cos(2\pi t)$ and no temporal white noise modes $\dot{w}_1, \dot{w}_2$  
of this tidal flow model,
is the  so-called cellular flow:
\begin{eqnarray}
\dot{x} & = &  \pi  \sin(\pi x)\cos(\pi y), \label{unperturb1} \\	             
\dot{y} & = & -\pi \cos(\pi x)\sin(\pi y). \label{unperturb2}
\end{eqnarray}
Figure \ref{cellular} shows the
phase portrait of this unperturbed flow 
(\ref{unperturb1})-(\ref{unperturb2}).

The partial differential equations for mean residence 
time $u$ of fluid particles
in a fluid domain $D$, and for the escape probability $p$
of fluid particles
cross a subboundary $\Gamma$ of $D$, are the following (in reversed time 
$s=T-t$), respectively,
\begin{eqnarray}
u_s & = &  \e (u_{xx} + u_{yy})     \nonumber  \\
     &  + & [\pi  \sin(\pi x)\cos(\pi y)+ \pi \lambda \cos(2\pi T-s) ] u_x 
     			\nonumber  \\
      &  - & \pi \cos(\pi x)\sin(\pi y)   u_y  +  1,   \\
p_s & = &   \e (p_{xx} + p_{yy})     \nonumber  \\
    & + & [\pi  \sin(\pi x)\cos(\pi y)+ \pi \lambda \cos(2\pi T-s) ]  p_x
    			\nonumber  \\
     & - & \pi \cos(\pi x)\sin(\pi y) p_y.               
\end{eqnarray}

In practical numerical simulations, the ``final" time 
$T$ should be taken big enough so that the solutions $u$, $p$
do not change within a reasonable tolerance. To do so, we monitor
the mean-square difference of $u$ (and also $p$) at time $T$ and
$T+1$. When this difference is within a  reasonable tolerance 
(we use $0.001$), we take the $T$ as the  ``final" time;
otherwise, we increase the value of $T$ and do the simulation again,
until the tolerance criterion is met.

\begin{figure}[tbp] 
\hbox to \hsize{\hfil \psfig{figure=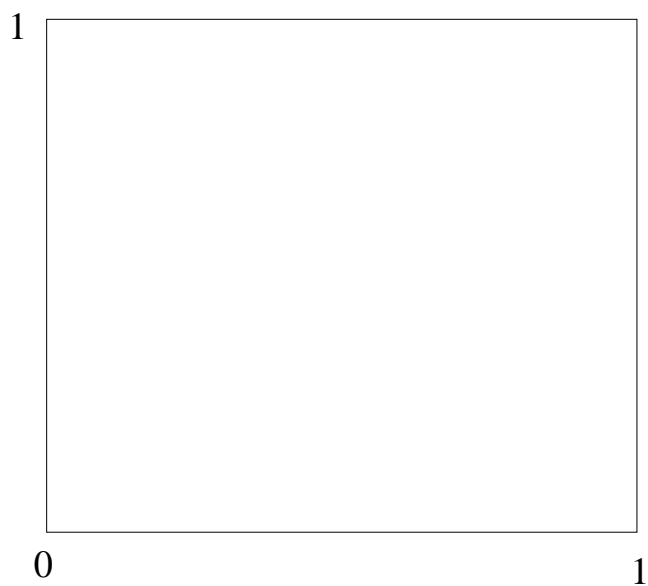,height=3in}\hfil}
\caption{A fluid domain $D$}
\label{cell}
\end{figure}

We take a fluid domain $D$ to be a typical cell, i.e., the unit square, in the
unperturbed flow; see Figure \ref{cell}.

\begin{figure}[tbp] 
\hbox to \hsize{\hfil \psfig{figure=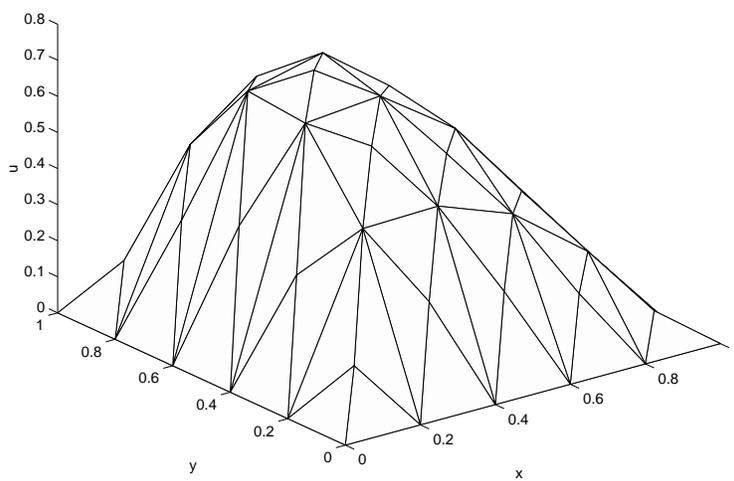,height=3in}\hfil}
\caption{Mean residence time after time $t=10$}
\label{t10}
\end{figure}

\begin{figure}[tbp] 
\hbox to \hsize{\hfil \psfig{figure=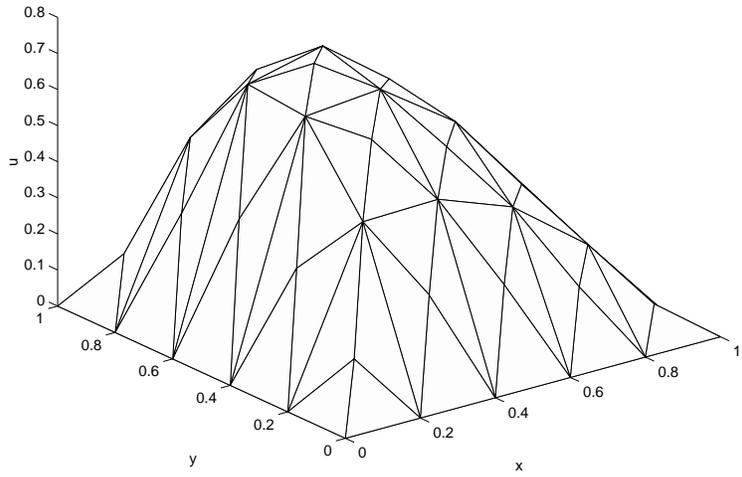,height=3in}\hfil}
\caption{Mean residence time after time $t=50.5$   }
\label{t60}
\end{figure}

\begin{figure}[tbp] 
\hbox to \hsize{\hfil \psfig{figure=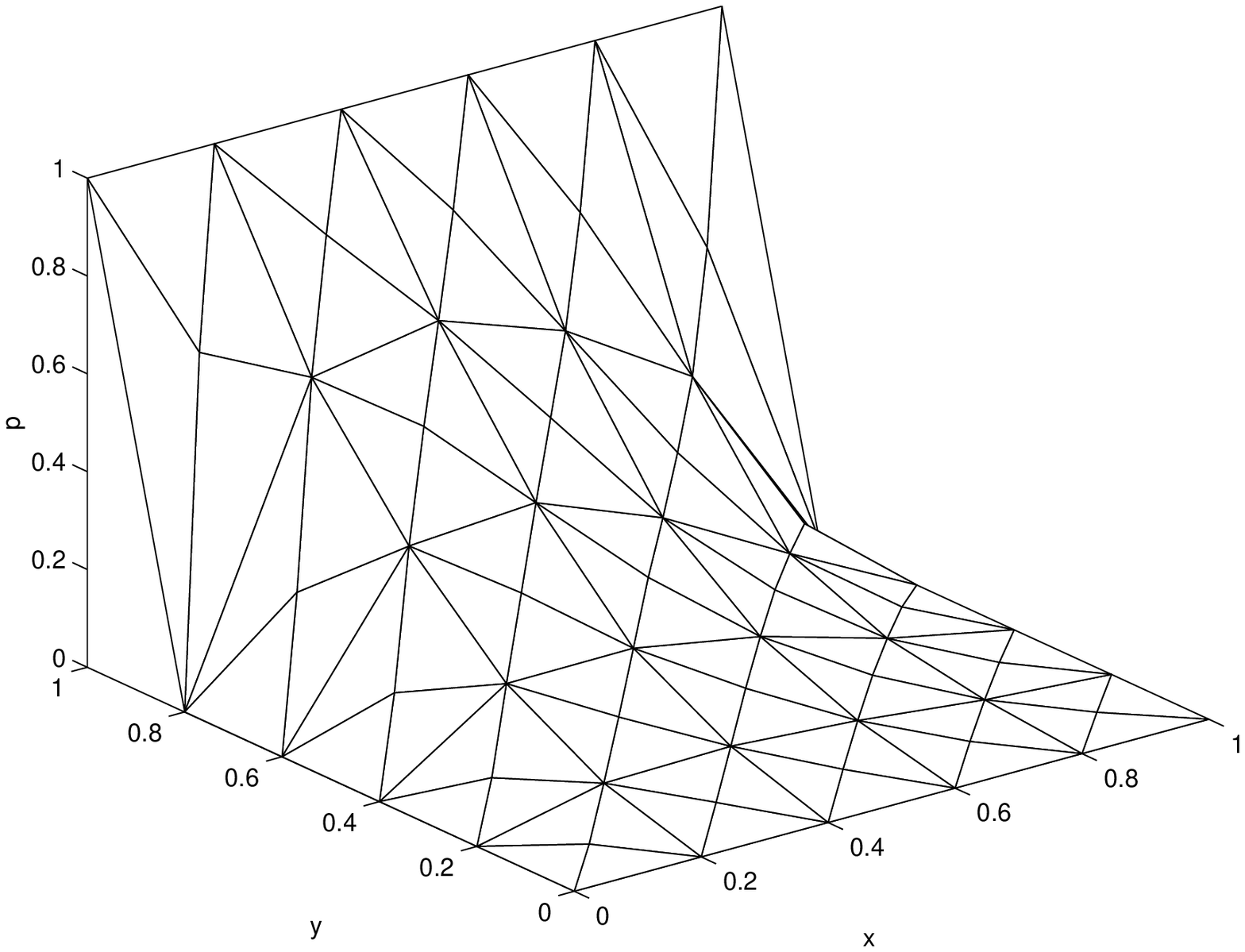,height=3in}\hfil}
\caption{Escape probability crossing the top boundary $y=1$ after time $t=10$}
\label{p10}
\end{figure}

\begin{figure}[tbp] 
\hbox to \hsize{\hfil \psfig{figure=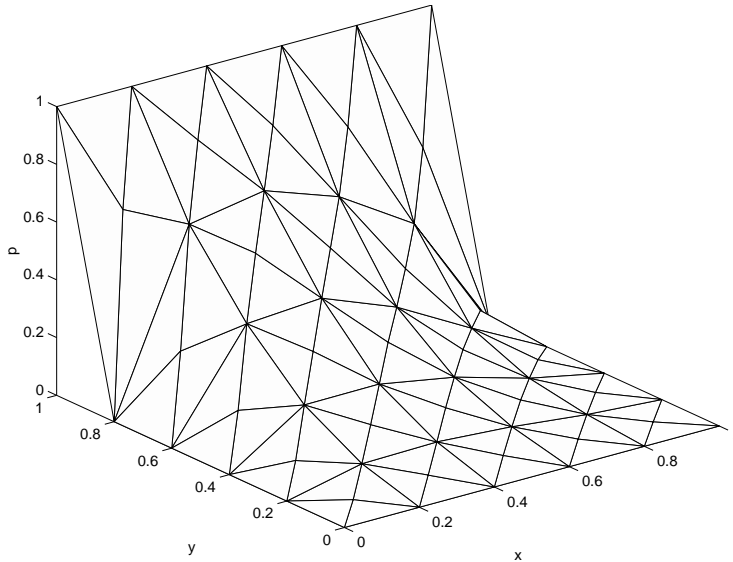,height=3in}\hfil}
\caption{Escape probability crossing the top boundary $y=1$ after time $t=60$}
\label{p60}
\end{figure}

Unlike the stochastic systems with steady drift as studied by 
Brannan, Duan and Ervin \cite{Brannan-exit}, the mean residence time 
and escape probability depend on time (although the change
is small in this specific example); see Figures \ref{t10},
\ref{t60}, \ref{p10} and \ref{p60}. All these plots are for $\lambda =1$
and $\e=0.1$.  
The mean residence time 
and escape probability for $0<\lambda <3$ and $0<\e<0.1$ 
display similar features.

In this flow model, it turns out that the average escape probability
crossing the top boundary $y=1$ does not change much with time, 
with value around $0.2639$. We observe similar features for crossing
other three side boundaries.

In the tidal flow model (\ref{tidal1})-(\ref{tidal2})  with  only
time-periodic ``tidal" mode  
$\pi  \lambda \cos(2\pi t)$,  
Beerens and Zimmerman \cite{Beerens_Zimmerman}  found that there
are ``islands" in the tidal flow   and the fluid particles
trapped in such islands will never escape, for $0< \lambda < 3$ as used
by Beerens and Zimmerman \cite{Beerens_Zimmerman}. This phenomenon
is common in non-dissipative, Hamiltonian planar  systems.
Although the dissipation in the oceans is small,
``no matter how small the dissipation is, the (oceanic)
fluid has substantial time to experience the action of dissipative forces"
\cite{Pedlosky2}.  So this ``islands" phenomenon does not appear
to be likely in realistic tidal flows; see more physical discussions
in \cite{Gill, Pedlosky, Salmon2}.
 
While in our tidal flow model (\ref{tidal3})-(\ref{tidal4})
with both time-periodic ``tidal" mode  
$\pi  \lambda \cos(2\pi t)$ and  temporal white noise modes, 
all fluid particles  will eventually escape from any fluid domain
in finite time after we first observe them; see 
Figures \ref{t10}, \ref{t60}. This feature is true for
any $\e >0$. It appears that our stochastic tidal model
is a little more realistic than Beerens 
and Zimmerman's model \cite{Beerens_Zimmerman}.  We remark again that
we use this simple tidal flow model
to demonstrate the applications of mean residence time and escape probability.

\bigskip
 
In summary, in this paper we have discussed  the quantification 
of fluid transport between flow
regimes of different characteristic motion by escape probability and mean
residence time,
 developed numerical  algorithms   to solve for
escape probability and mean residence time, and  
  applied these
ideas and numerical   algorithms   to 
a tidal flow model.

\bigskip
 {\bf Acknowledgement.} 
 
We would like to thank Ludwig Arnold for
very useful  suggestions.  
We would also thank the hospitality of
the Oberwolfach  Mathematical  Research
Institute, Germany. This work was supported by the NSF Grant DMS-9704345.

\end{document}